\documentclass{cta-author}
\usepackage{mathtools}
\usepackage{mathabx}
\usepackage{float}
\usepackage{balance}
\usepackage{caption}
\usepackage{multirow} 
\usepackage{subcaption} 
\usepackage{xcolor}

\usepackage{booktabs,tabularx,graphicx}
\usepackage{amsmath}
{}
{}
{}
\usepackage{wrapfig}
\begin{document}

\supertitle{Submission Template for IET Research Journal Papers}

\title{Intelligent Protection \& Classification of Transients in Two-Core Symmetric Phase Angle Regulating Transformers}

\author{\au{Pallav Kumar Bera$^{1\corr}$}, \au{Can Isik$^{1}$}}
\address{\add{1}{Electrical Engineering and Computer Science , Syracuse  University, Syracuse, New York, USA}
\email{pkbera@syr.edu}}

\begin{abstract}

{This paper investigates the applicability of time and time-frequency features based classifiers to distinguish internal faults and other transients -- magnetizing inrush, sympathetic inrush, external faults with current transformer saturation, and overexcitation -- for Indirect Symmetrical Phase Angle Regulating Transformers (ISPAR). Then the faulty transformer unit (series/exciting) of the ISPAR is located, or else the transient disturbance is identified. An event detector detects variation in differential currents and registers one-cycle of 3-phase post transient samples which are used to extract the time and time-frequency features for training seven classifiers. Three different sets of features -- wavelet coefficients, time-domain features, and combination of time and wavelet energy -- obtained from exhaustive search using Decision Tree, random forest feature selection, and maximum Relevance Minimum Redundancy are used. The internal fault is detected with a balanced accuracy($\bar{\eta}$) of 99.9\%, the faulty unit is localized with $\bar{\eta}$ of 98.7\% and the no-fault transients are classified with $\bar{\eta}$ of 99.5\%. The results show potential for accurate internal fault detection and localization, and transient identification. The proposed scheme can supervise the operation of existing microprocessor-based differential relays resulting in higher stability and dependability. The ISPAR is modeled and the transients are simulated in PSCAD/EMTDC by varying several parameters.}
\end{abstract}

\maketitle

\section{Introduction}\label{sec1}
Phase Angle Regulating Transformers or Phase Angle Regulators (PARs) are used to control real power flow in networked power systems and make certain that the ratings of transmission equipment are not exceeded during contingency conditions. The performance of a PAR affects the continuous and stable operation of a power system. Having a lower successful operating rate than the transmission lines a transformer protection system is challenged under various operating conditions. Internal faults are electrically detected in a transformer mainly with differential relays, overcurrent relays, and ground fault relays. Differential relays detect and clear faults faster and locate them accurately. Electromechanical, solid-state, analog, and microprocessor-based relays are usually used as differential relays. Differential protection is also used for different kinds of transformers including PARs and it's operation highly depends on appropriate analysis of the different electromagnetic transient events.

Differential protection being the foremost, however, suffers from unwanted tripping in case of magnetizing inrush, {current transformer (CT)} saturation during external faults and overexcitation conditions. These problems are addressed by current based methods in two ways: using harmonics restraint and waveshape identification methods \cite{waveshape}. The changing complexity and operating modes of power system e.g with the increase in nonlinear loads the stability of these methods are threatened. Percentage differential relay with restraint actuated by restraining current and/or harmonic components of operating current is generally used in current differential schemes. The second harmonic component of operating current identifies inrush and fifth identifies overexcitation. The traditional methods of second harmonic restraint method \cite{harmonic} to detect magnetizing inrush fail because of lower second harmonics content in certain cases (modern core transformer materials \cite{modern_core}, presence of higher residual magnetism). Moreover, the sensitivity of the protection is compromised in case of internal faults due to higher second harmonics during internal faults with CT saturation or the presence of distributed and series compensation capacitance \cite{tx3pro}. The fifth harmonic restraint may fail in case of internal faults during overexcitation. Use of fourth harmonic with second in case of inrush and adaptive fifth harmonic pick up in case of overexcitation improves security. Harmonic blocking and harmonic sharing (modified harmonic blocking) are also used in some cases. An improved algorithm based on dwell-time principle (intervals of flat and low currents in inrush current) is used in digital differential relays. Modern relays also have options of even-harmonic restraint /blocking to dc blocking and fifth harmonic restraint mechanisms for higher security, yet the challenges exist. CT saturation during external faults may also cause false trips due to the inefficient setting of commonly used dual-slope biased differential relays \cite{ctsat1}. Differential relays also fail to detect ground faults near neutral of ground wye-connected transformer winding or ground faults for system grounded with an impedance. Restricted earth fault schemes can be used to detect such faults \cite{transformerguide}. Intelligent techniques can be used to identify such transient conditions where the tripping is unwanted and optimize the operation of differential relays.

PARs can be categorized based on the number of 3-phase(ph) magnetic cores and the magnitude of source end voltage with respect to the load end. Two-Core Symmetric Phase Angle Regulating Transformer or Indirect Symmetrical Phase Angle Regulators have the same source and load end voltages and two 3-phase magnetic core units: series and exciting (Fig. \ref{par23}c). They are conventionally used PARs with higher security against high voltage levels. To regulate power flow the exciting unit creates the required phase difference through the load tap changer (LTC) connections and the forward/backward transition can be achieved in the series secondary using an advance-retard-switch (ARS) or in the exciting secondary using an ARS or LTC change-over selectors \cite{ibrahim}.

Taking into account the high repair and replacement cost and to limit further damages, the PARs require a sensitive, secure, and dependable protection system. Maintaining dependability for in-zone faults and security against no-fault conditions is a challenge for protection systems of PARs. High sensitivity is required to detect turn to turn and ground faults. Also, methods used to compensate differential relays in the case of traditional transformers with a fixed phase shift are not applicable in PARs with variable phase shift\cite{pstguide}. Consequently, special consideration is required when designing the protection system. 

Various intelligent methods have been proposed by different authors to distinguish internal faults and magnetizing inrush in Power Transformers. Neural Networks (NN) and spectral energies of the wavelet components are used to discriminate internal faults and inrush in \cite{ann1}. Support Vector Machines (SVM) and Decision Tree based transformer protection are proposed in \cite{SVM1,SVM2} and \cite{dt1,dt2,dtwt}.
Probabilistic Neural Network (PNN) has been used to detect different conditions in Power Transformer operation in \cite{tripathyrb}. Random Forest (RF) is used in \cite{Shahrfc} to discriminate internal faults and inrush currents. Although several studies have been conducted with Power transformers, this problem is still insufficiently explored in PARs.
Few literatures demonstrate the discrimination of faults and other disturbances in PARs. The internal faults are distinguished from magnetizing inrush using WT and classified using NN in \cite{pallav}. In \cite{pallav2} different internal faults in series and exciting transformers of the ISPAR  are classified using RFC. 
These studies cannot be considered as conclusive because sympathetic inrush, CT saturation during external faults and overexcitation conditions have not been considered. The problem is not studied consistently because a complete model of PAR with provision to simulate internal faults is absent. To properly address this question, ISPAR is modeled and internal faults are detected and transients are classified in an interconnected system for Phase Angle Regulators (PAR) and Power Transformers \cite{systempallav}.

This paper studies the applicability of time and time-frequency domain features to discriminate the internal faults and other transient disturbances including magnetizing inrush, CT saturation during external faults, overexcitation for a Phase Angle Regulator. The ISPAR is modeled in PSCAD using two-winding and three-winding transformers to simulate the internal faults. A series of time and wavelet features are extracted and then selected using feature selection algorithms. Seven classifiers trained on 60,552 transient cases simulated in PSCAD/EMTDC by varying various system parameters demonstrate the validity of the proposed scheme.

The rest of the paper is organized as follows. Section 2 describes the modeling and simulation of the internal faults and the other transient events in the ISPAR. Section 3 presents the proposed differential protection scheme which includes the detection of an event, feature extraction and selection, and introduction of the classifiers used. Section 4 consists of the performance of different classifiers on the features selected in the previous section for the detection of internal faults, identification of faulty unit and transient disturbances. The last section concludes the paper. 

\section{Modeling and Simulation}\label{sec2}
\begin{figure}[ht]
\centerline{\includegraphics[width=3.0 in, height= 0.8 in]{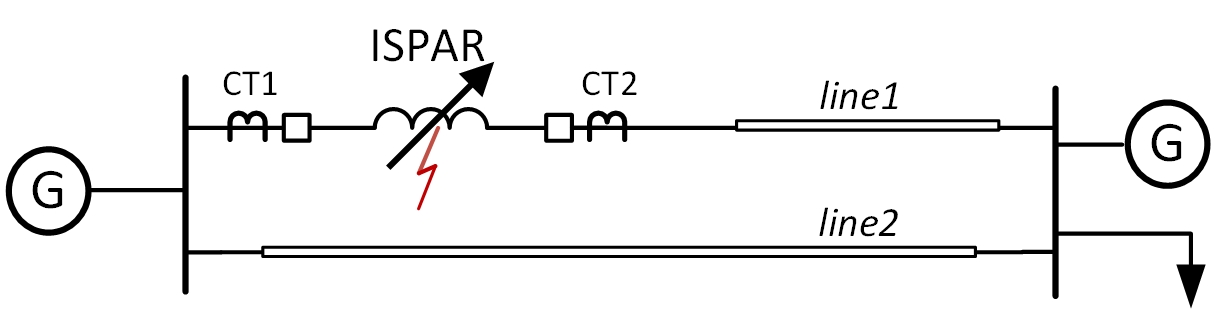}}
\caption{Single line diagram of the PSCAD model consisting of the AC sources, transmission lines, ISPAR, CTs, and 3-phase load. }
\label{ckt}
\end{figure}

PSCAD/EMTDC is used for the modeling and simulation of the electromagnetic transients in the ISPAR. Figure \ref{ckt} shows the single line diagram of the model consisting of the AC sources, transmission lines, ISPAR, and 3-phase loads working at 60Hz. The ISPAR is rated at 500 MVA, 230kV/230kV, with phase angle variations of $\pm25^{\circ}$.

\begin{figure}[ht]
\centerline{\includegraphics[width=2.6 in, height= 2.2 in]{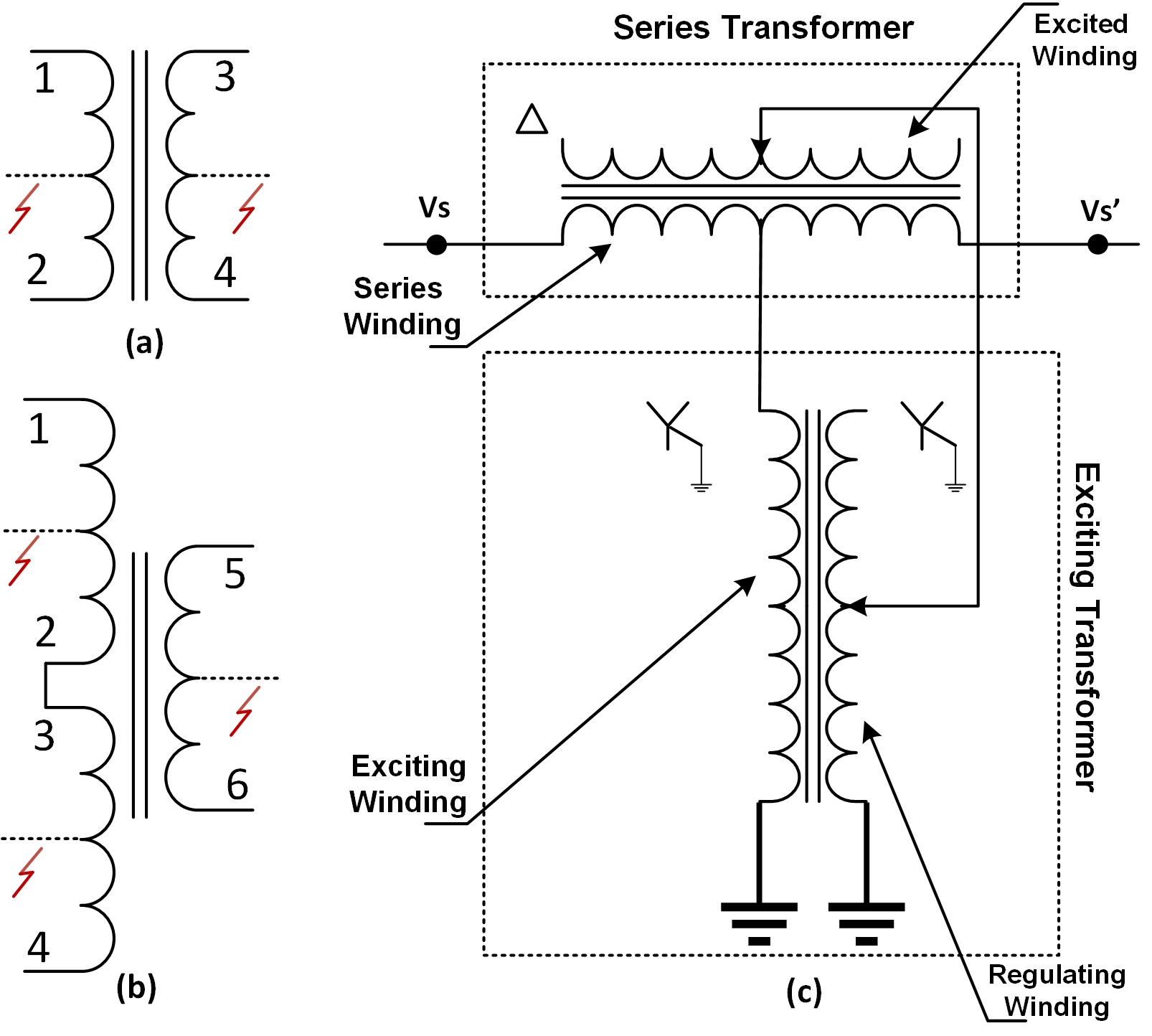}}
\caption{Developed (a) two-winding fault model, (b) three-winding fault model, and (c) Indirect Symmetrical Phase Angle Regulating Transformer model in PSCAD }
\label{par23}
\end{figure}

The fault model of ISPAR is not available in most simulation software. The single-phase two-winding transformer fault model (Figure \ref{par23}a) required for the exciting unit of ISPAR and the three-winding transformer fault model (Figure \ref{par23}b) required for the series unit of ISPAR for the simulating various internal faults were developed in PSCAD/ EMTDC \cite{systempallav}. The self-inductance terms \textit{Li} and the mutual inductance terms \textit{Mij} of the $6\times 6$ matrix in the voltage-current equation (Eqn.\ref{matrix}) of the three-winding transformer and $4\times 4$ matrix of the two-winding transformer are computed from voltage ratios, the inductive component of the no-load excitation currents ($I_m$) and the short-circuit tests. The modeled components have the provision to change the saturation characteristics, \% of winding faulted and other parameters. The Fortran script of the single-phase three-winding transformer is provided in the Appendix Section.
\begin{equation}\label{matrix}
\footnotesize
\begingroup 
\setlength\arraycolsep{1.7pt}
\begin{bmatrix}V1 \\ V2\\ V3\\V4\\V5\\V6
\end{bmatrix}=
\begin{bmatrix}
L1 & M12 & M13& M14&M15&M16\\
M21 & L2 & M23& M24&M25&M26\\
M31 & M32 & L3& M34&M35&M36\\
M41 & M42 & M43& L4&M45&M46\\
M51 & M52 & M53& M54&L5&M56\\
M61 & M62 & M63& M64&M65&L6\\
\end{bmatrix}
\cdot\dfrac{d}{dt}
\begin{bmatrix}I1 \\ I2\\I3\\I4\\I5\\I6
\end{bmatrix}
\endgroup
\end{equation}

In this work, the internal faults, magnetizing inrush and sympathetic inrush, external faults with CT saturation, and overexcitation transients for ISPAR are taken into account. In the following sections, these conditions are considered successively. The simulation run-time and fault duration time are 10.2 secs, and 0.05 secs (3 cycles) respectively in all the cases. The multi-run component is used to alter the parameter values to get the different simulation runs.

\subsection{Internal Faults}
The internal faults are created in the ISPAR series and exciting units. They include the faults occurring inside the enclosure and inside the CT locations.
The 46,872 fault cases which include basic internal faults (short circuits and phase (ph) faults), turn-to-turn, and winding-to-winding faults are simulated by varying the fault resistance, \% of winding under fault, fault inception time, forward or backward shift, and the LTC in the exciting unit. They are usually caused by the breakdown of insulation and require faster action by protective relays to limit the extent of damage.

\subsubsection{ Internal phase \& ground faults}
Winding ph-g faults (a-g, b-g, c-g), winding ph-ph-g faults (ab-g, ac-g, bc-g), winding ph-ph faults (ab, ac, bc), 3-ph and 3-ph-g faults are simulated in the primary (P) and secondary (S) sides of exciting and series transformer units in the ISPAR. Table \ref{tabintrnalfault} shows the values of different parameters of the ISPAR which are varied to get the instances for training and testing. In total, 33,264 internal phase and ground faults are simulated in the ISPAR.
\begin{table}[htbp]
\footnotesize
\renewcommand{\arraystretch}{0.8}
\setlength{\tabcolsep}{4pt}
    \centering
    \caption{Parameters for internal  phase  \&  ground  faults}
    \label{tabintrnalfault}
    \begin{tabular}{|l|l|}
\toprule
\textit{Variables}  & \textit{Values} \\ \midrule
\textit{fault resistance} &  0.01, 0.1 \&   1 $\Omega$     (3)   \\
\textit{\% of winding shorted} & 20\%, 50\%, 70\%  (3)\\
\textit{fault type }     & lg, llg, ll, lll \& lllg in 3 phs  (11) \\
\textit{fault inception time }      & 10s to 10.0153s in steps of 1.38ms (12)\\
\textit{faulty unit \& phase }          &     \begin{tabular}{@{\extracolsep{\fill}}l}ISPAR Exciting unit (P \& S) (2)\\ \& ISPAR Series unit (P \& S) (2) \end{tabular}  \\
\textit{phase shift} & forward and backward  (2)\\ 
\textit{LTC}  & 0.2,0.4,0.6,0.8,1[1 \& 0.5 in ISPAR exciting] \\ \midrule
\multicolumn{2}{|l|}{}                                                \\ 
\multicolumn{2}{|l|}{\multirow{-2}{*}{{ \begin{tabular}[c]{@{}l@{}}  Series unit cases = $3\footnotesize{\times}3\footnotesize{\times}11\footnotesize{\times}12\footnotesize{\times}2\footnotesize{\times}2\footnotesize{\times}5$ = 23760    \\  
Exciting unit cases = $3\footnotesize{\times}3\footnotesize{\times}11\footnotesize{\times}12\footnotesize{\times}2\footnotesize{\times}2\footnotesize{\times}2$ = 9504\end{tabular}}}} \\ \hline
\end{tabular}
\end{table}

\subsubsection{Turn-to-turn (T-T) faults} Turn-to-turn insulation failures are responsible for a major percentage of faults in any transformer. Thermal, mechanical and electrical stresses degrade the insulation and cause turn-to-turn faults which may lead to more serious faults and interwinding faults if not detected quickly \cite{turn}. They are challenging to monitor and detect, particularly when \% turns shorted is lower. Table \ref{tab_ww_tt} shows the values of different parameters of the series and exciting unit of ISPAR used to simulate 9,072 turn-to-turn faults.

\begin{figure}[ht]
\centerline{\includegraphics[width=3.2 in, height= 2.2 in]{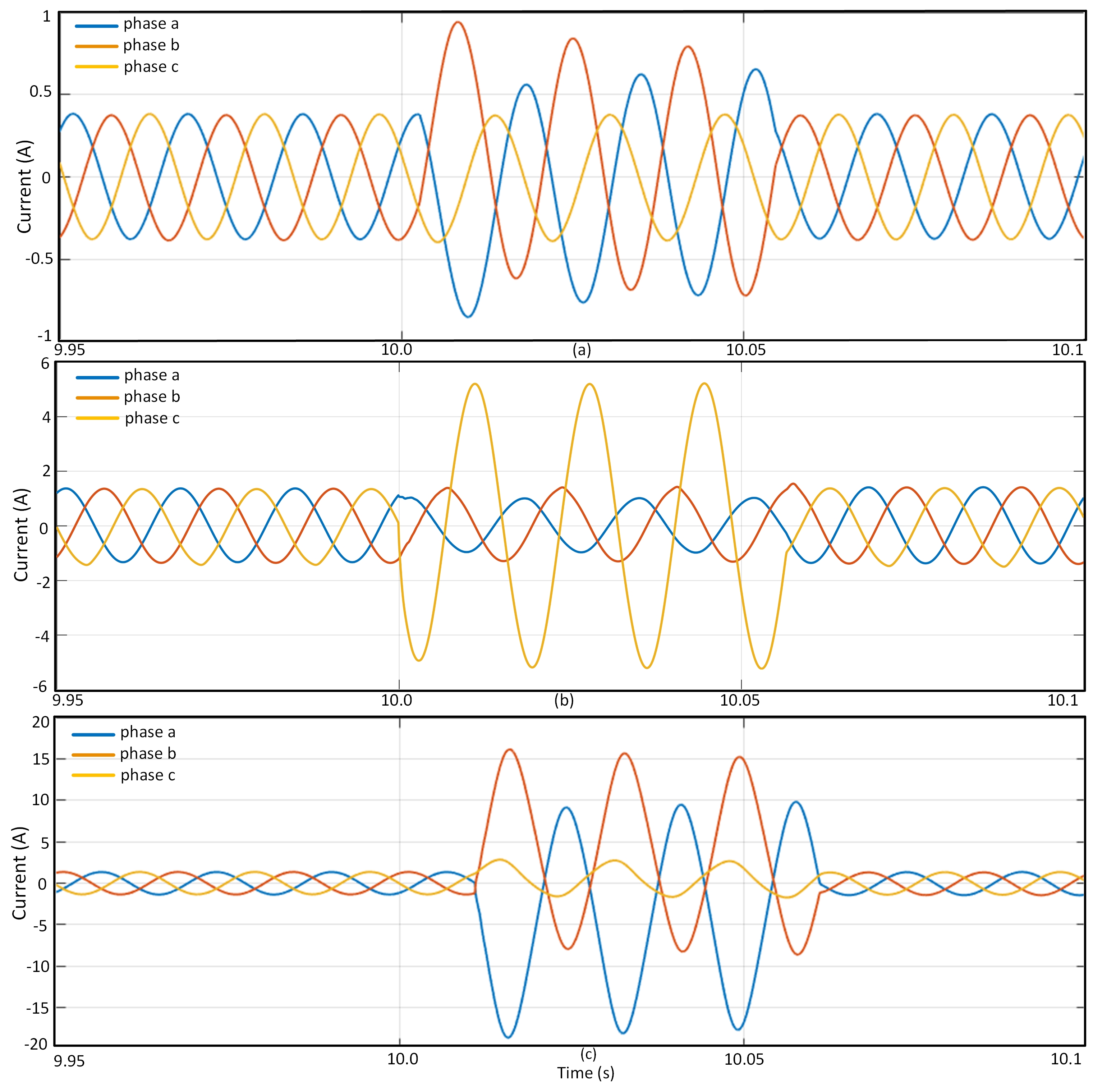}}
\vspace{0mm}\caption{3-phase differential currents for (a) turn-to-turn fault in  primary of series unit (b) ph-g fault in primary of exciting unit, (c) ph-ph fault in secondary of exciting unit.}
\label{tt}
\end{figure}

\begin{table}[ht]
\centering
\renewcommand{\arraystretch}{0.8}
\footnotesize
\setlength{\tabcolsep}{4pt}
    \centering
    \caption{Parameters for winding-to-winding \& turn-to-turn fault}
    \label{tab_ww_tt}
    \begin{tabular}{|l|l|}
\toprule
\textit{Variables}  & \textit{Values} \\ \midrule
\textit{fault resistance} &  0.01, 0.5 \&   1 $\Omega$     (3)   \\
\textit{\% of winding shorted} & 20\%, 50\%, 70\% (3)\\
\textit{fault inception time}       & 10s to 10.0153s in steps of 1.38ms (12)\\
\textit{faulty unit \& phase}          &     \begin{tabular}{@{\extracolsep{\fill}}l} ISPAR Exciting ph A,B,C (P \& S) (6)\\ \& ISPAR Series ph A,B,C (P \& S) (6) \end{tabular}  \\
\textit{phase shift} & forward and backward  (2)\\ 
\textit{LTC}  & 0.2,0.4,0.6,0.8,1 [1 \& 0.5 in ISPAR exciting] \\ \midrule
\multicolumn{2}{|l|}{}                                                                                                                                                                                                                                                             \\
\multicolumn{2}{|l|}{}                                                                                                                                                                                                                                                                                  \\
\multicolumn{2}{|l|}{}                                                                                                                                                                                                \\
\multicolumn{2}{|l|}{\multirow{-4}{*}{{ \begin{tabular}[c]{@{}l@{}}\footnotesize Series unit(T-T) cases = $3\footnotesize{\times}3\footnotesize{\times}12\footnotesize{\times}6\footnotesize{\times}2\footnotesize{\times}5$ = 6480\\   Exciting unit(T-T)   cases = $3\footnotesize{\times}3\footnotesize{\times}12\footnotesize{\times}6\footnotesize{\times}2\footnotesize{\times}2$ = 2592 \\  Series unit(W-W) cases = $3\footnotesize{\times}3\footnotesize{\times}12\footnotesize{\times}3\footnotesize{\times}2\footnotesize{\times}5$ = 3240\\   Exciting unit(W-W) cases = $3\footnotesize{\times}3\footnotesize{\times}12\footnotesize{\times}3\footnotesize{\times}2\footnotesize{\times}2$ = 1296\end{tabular}}}} \\ \hline
\end{tabular}
\end{table}

\subsubsection{Winding-to-winding (W-W) faults}
The electrical and thermal stresses due to short circuits and transformer aging also reduce the mechanical and dielectric strength of the winding and result in degradation of the insulation between LV and HV winding and may damage the winding eventually \cite{turn}. Table \ref{tab_ww_tt} shows the values of different parameters of the series and exciting unit of ISPAR used to simulate 4,536 winding-to-winding faults.

Fig.\ref {tt}(a) shows the differential currents for a turn-to-turn fault in primary of series unit with LTC = 0.4, fault inception time = 10.00276s, phase shift = forward, fault resistance = 0.01$\Omega$, and \% turns shorted = 20; Fig.\ref {tt}(b) shows the differential currents for a ph-g fault (a-g) in primary of exciting unit with LTC = half, fault inception time = 10.0s, phase shift = backward, fault resistance = 1.0$\Omega$ and \% turns shorted = 70; and Fig.\ref {tt}(c) shows the differential currents for a ph-ph fault (bc) in secondary of exciting unit with LTC = half, fault inception time = 10.01104s, phase shift = forward, fault resistance = 0.1$\Omega$ and \% turns shorted = 50.

\subsection{Overexcitation}
Faults due to over fluxing develop slowly and cause deterioration of insulation and may lead to major faults. Overexcitation causes heating, and vibration and can damage the transformer. Since it is difficult for differential protection to control the amount of overexcitation any transformer can tolerate, the tripping of the differential element during overexcitation is undesirable. Generally, 5th harmonic restraint is used to restrain the operation of differential relays \cite{5th}. Several operations may lead to overexcitation in electrical systems. Here, two such situations have been modeled : overvoltage during load rejection and capacitor switching.

\begin{table}[ht]
\renewcommand{\arraystretch}{0.8}
\setlength{\tabcolsep}{7pt}
\footnotesize
\centering
\caption{Parameters for Overexcitation}\label{tabover}
\begin{tabular}{|l|l|}\toprule
\textit{Variables}  & \textit{Values} \\ \midrule
\textit{switching} & load(3) \& capacitor(3)\\
\textit{switching time  }     & 10s to 10.0153s in steps of 1.38ms   (12)     \\
\textit{LTC}  & 0.2 to full tap in steps of 0.2 (5)                       \\
\textit{phase shift} & forward and backward (2)\\ \midrule
\multicolumn{2}{|l|}{Total = $6\footnotesize{\times}12\footnotesize{\times}5\footnotesize{\times}2$ = 720}\\
\hline
\end{tabular}{}
\end{table}

\subsection{Magnetizing inrush} 
\begin{figure}[ht]
\centerline{\includegraphics[width=1.8 in, height= 1.7 in]{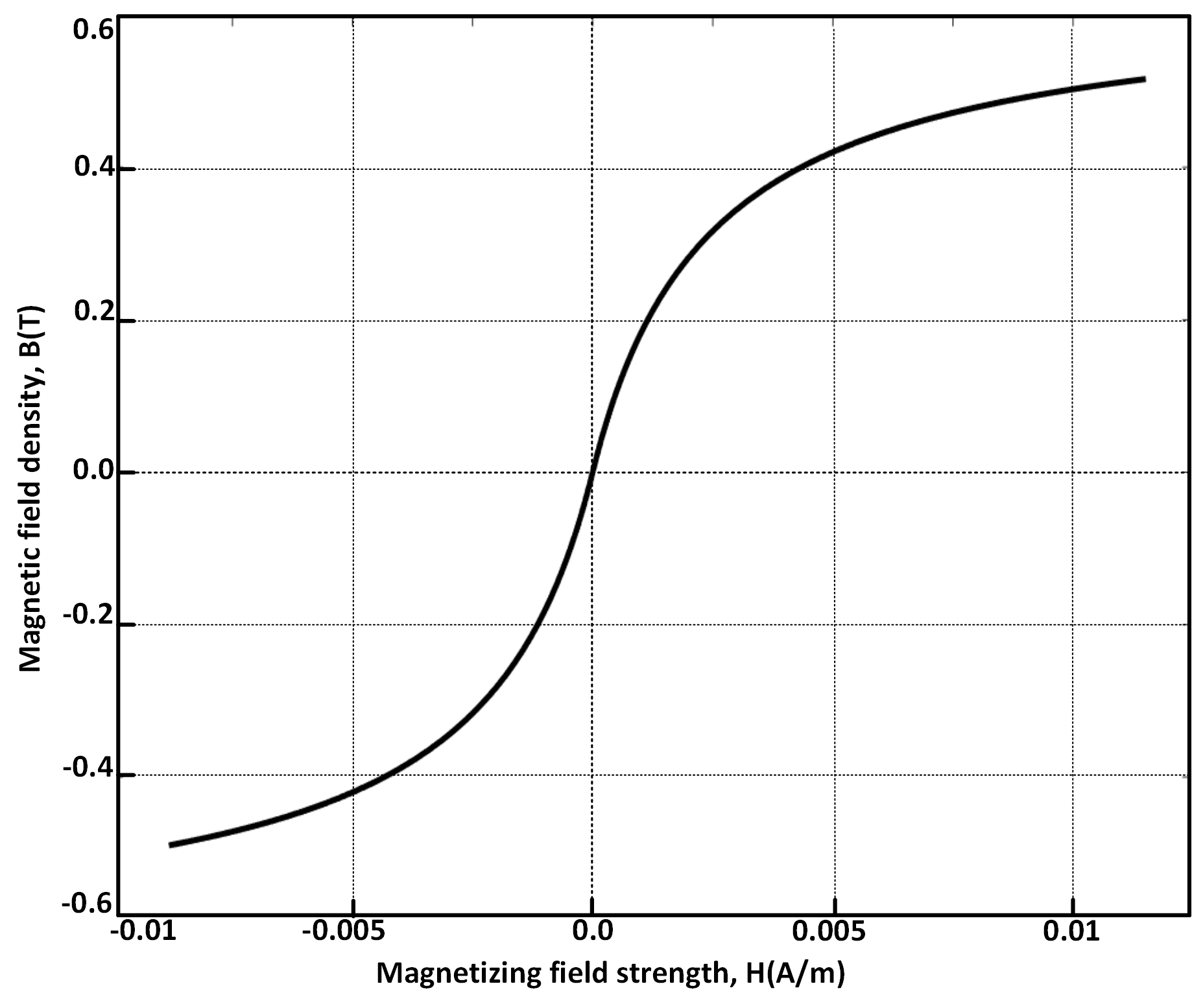}}
\vspace{0mm}\caption{B-H curve of transformer core in PSCAD/EMTDC}
\label{bh}
\end{figure}

\begin{figure}[ht]
\centerline{\includegraphics[width=3.2 in, height= 1.6 in]{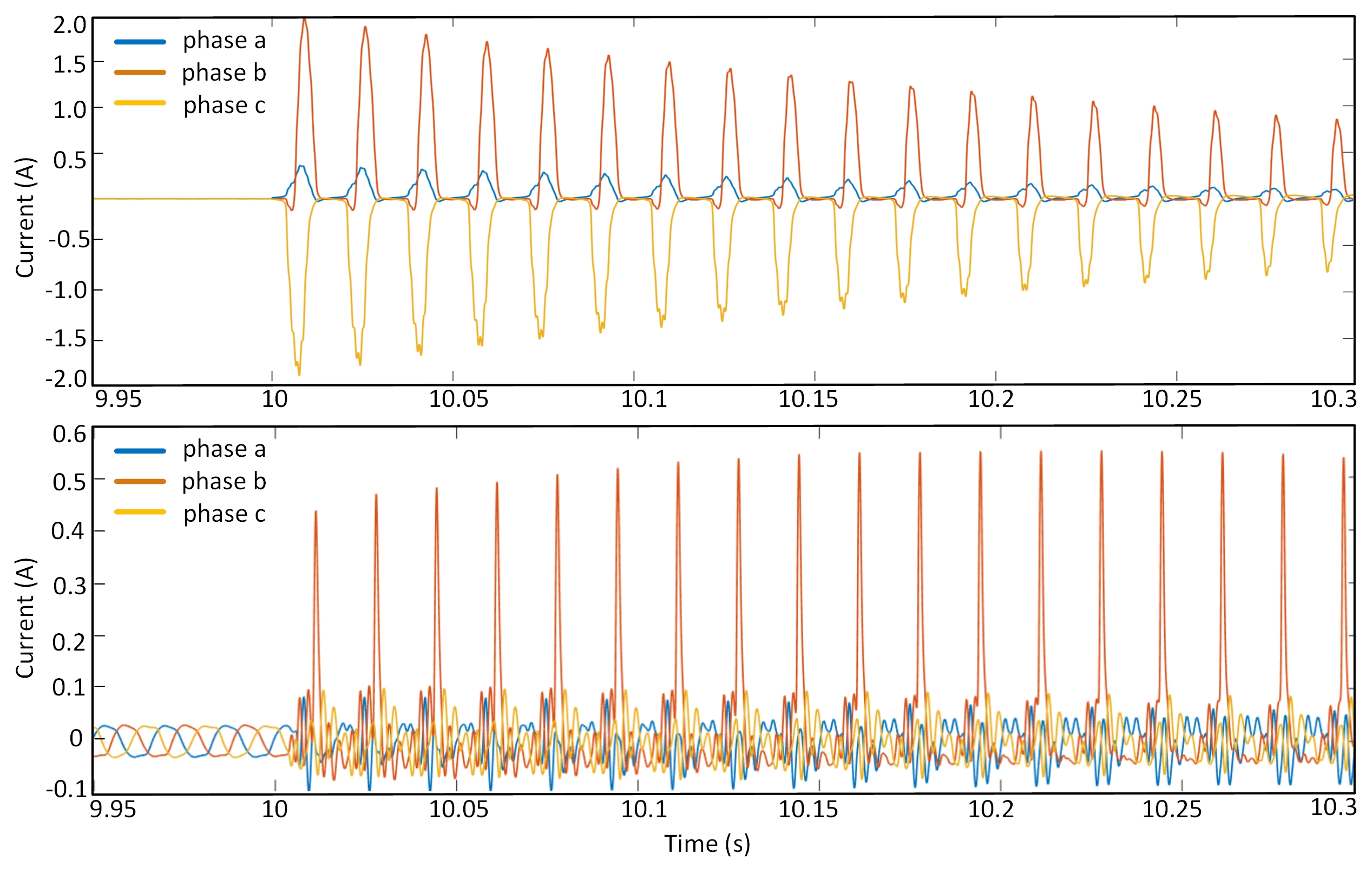}}
\vspace{0mm}\caption{3-phase differential currents for (a) Magnetizing inrush, and (b) Sympathetic inrush}
\label{ext}
\end{figure}

When a transformer is energized, an inrush current of the order of 8-10 times of normal current flows because of the saturation of the transformer core. Since the high current flows only on one side of the transformer the differential relay mal-operates. The second harmonic content of inrush currents is claimed to be more than 15\% \cite{15per}. But harmonic restraint relays may fail to detect inrush currents in some special scenarios and in case of transformers with modern core materials. The flux in a transformer core is given by:
\begin{equation}
\Phi = \Phi_R + \Phi_m cos\omega T - \Phi_m cos\omega( t + t')
\end{equation} where, $\Phi_R$=residual flux,  $\Phi_m$ = maximum flux, t$'$ = switching time. Here, $\Phi_R$ and t$'$ which influence the magnitude and duration of inrush currents are taken as variables. DC sources are used to get the desired residual fluxes in the single-phase two-winding transformers. The accurate values for the DC currents in phase-A, B, and C are obtained from the B-H curve of the transformer core material as shown in Figure \ref{bh}. Table \ref{inrush} shows the values of different parameters used to get the training and testing data for magnetizing inrush. Figure \ref {ext}(a) shows typical differential inrush currents in the 3-phases for tap=full, switching time =10s, phase shift = forward, and remnant flux density = 0 in all three phases. The exciting transformer unit in the ISPAR is considered to be responsible for the magnetizing inrush currents \cite{pstguide}.

\begin{table}[ht]
\renewcommand{\arraystretch}{0.8}
\footnotesize
\centering
\caption{Parameters for Magnetizing and Sympathetic inrush}\label{inrush}
\begin{tabular}{|l|l|}\toprule
\textit{Variables}  & \textit{Values} \\ \midrule
\textit{residual flux} & $\pm80\%, \pm60\%, \pm40\%, 0\% $ in 3 phs; $7\footnotesize{\times}3$=(21)\\
\textit{switching time }      & 10s to 10.0153s in steps of 1.38ms   (12)     \\
\textit{LTC}  & 0.2 to full tap in steps of 0.2 (5)                       \\
\textit{phase shift} & forward and backward (2)\\ \midrule
\multicolumn{2}{|l|}{Total = $21\footnotesize{\times}12\footnotesize{\times}5\footnotesize{\times}2$ = 2520}\\
\hline
\end{tabular}{}
\end{table}

\subsection{Sympathetic Inrush} Sympathetic inrush can occur in an in-service ISPAR when another incoming transformer is energized in series or parallel in a resistive power network. The asymmetrical flux change per cycle which drives the ISPAR to saturation is given by:
\begin{equation}\Delta\Phi=\int_{t}^{2\pi + t}[(R_{sys} + R_{PAR})i_1 + R_{sys}\cdot i_2]
\end{equation} where $R_{sys}$ = system resistance , $R_{PAR}$ = resistance of ISPAR, $i_1$ and $i_2$ are magnetizing currents of ISPAR and the incoming transformer.  This interaction between the incoming transformer and the ISPAR may lead to failure of differential relays and cause prolonged harmonic over-voltages \cite{sym2}. Use of superconducting winding, soft magnetic material in the core, and CT partial transient saturation are some factors that cause the mal-operations \cite{modern_core} \cite{ctsat}. The magnitude and direction of the residual flux of the incoming transformer, switching time, and system resistance have a significant influence on the magnitude of inrush current \cite{kumbhar}. In this work, the magnitude and direction of residual flux and the switching time are altered and the incoming transformer is connected in parallel. Table \ref{inrush} shows the values of different parameters used to simulate different cases for sympathetic inrush.
Figure  \ref {ext}(b) shows the 3-phase differential current for tap=0.4, switching time =10s, phase shift = forward and no residual flux.

\subsection{External fault with CT saturation}
\begin{figure}[ht]
\centerline{\includegraphics[width=3.2 in, height= 2.3 in]{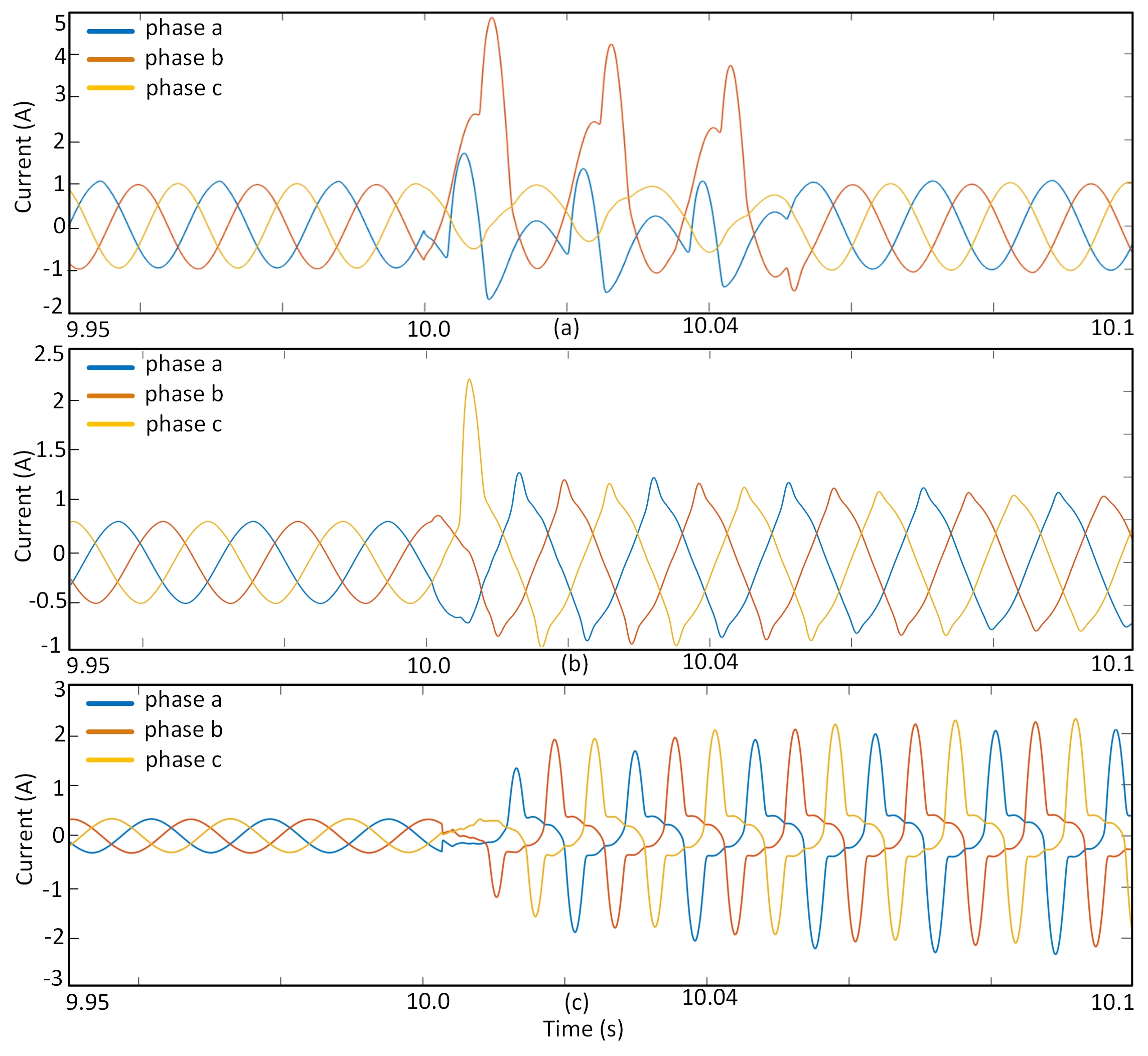}}
\vspace{0mm}\caption{3-phase differential currents for (a) CT saturation during external faults, (b) overexcitation due to load rejection, and (c) overexcitation due to capacitor switching. }
\label{tran}
\end{figure}

External short circuits constitute external faults. They stress the ISPAR and reduce the transformer life. The differential current becomes non zero due to dissimilar saturation in the CTs on the two sides of the ISPAR in case of heavy through faults and may lead to a false trip. Raising the bias threshold might ensure stability (i.e. no mal-operation), but the sensitivity for in-zone internal faults gets reduced. The effect of CT saturation is double-edged and the percentage restraint does not address security and dependability simultaneously. 
The external faults with CT saturation are simulated on the line1 and line2.
The values for the different parameters are given in table \ref{externaltab}. Figure  \ref {tran}(a) shows the 3-phase differential current for an external lg fault when tap=0.2, phase shift =forward, inception time =10s, and fault resistance = 0.01$\Omega$.

\begin{table}[ht]
\renewcommand{\arraystretch}{0.8}
\footnotesize
\centering
\caption{Parameters for External faults on \textit{line1 \& line2}}
\label{externaltab}
\begin{tabular}{|l|l|}\toprule
\textit{Variables } & \textit{Values} \\ \midrule
\textit{fault resistance} & 0.01, 0.1 \&   1 $\Omega$     (3)   \\
\textit{fault type }     & lg, llg, ll, lll \& lllg in 3 phs  (11) \\
\textit{fault inception time  }     & 10s to 10.0153s in steps of 1.38ms    (12)    \\
\textit{LTC}  & 0.2 to full tap in steps of 0.2 (5)\\
\textit{phase shift} & forward and backward  (2)\\ 
\textit{fault location }& line1 \& line2 (2)\\ \midrule
\multicolumn{2}{|l|}{Total = $3\footnotesize{\times}11\footnotesize{\times}12\footnotesize{\times}5\footnotesize{\times}2\footnotesize{\times}2$ = 7920}\\
\hline
\end{tabular}{}
\end{table}

\begin{figure}[ht]
\centerline{\includegraphics[width=3.2 in, height= 3.3 in]{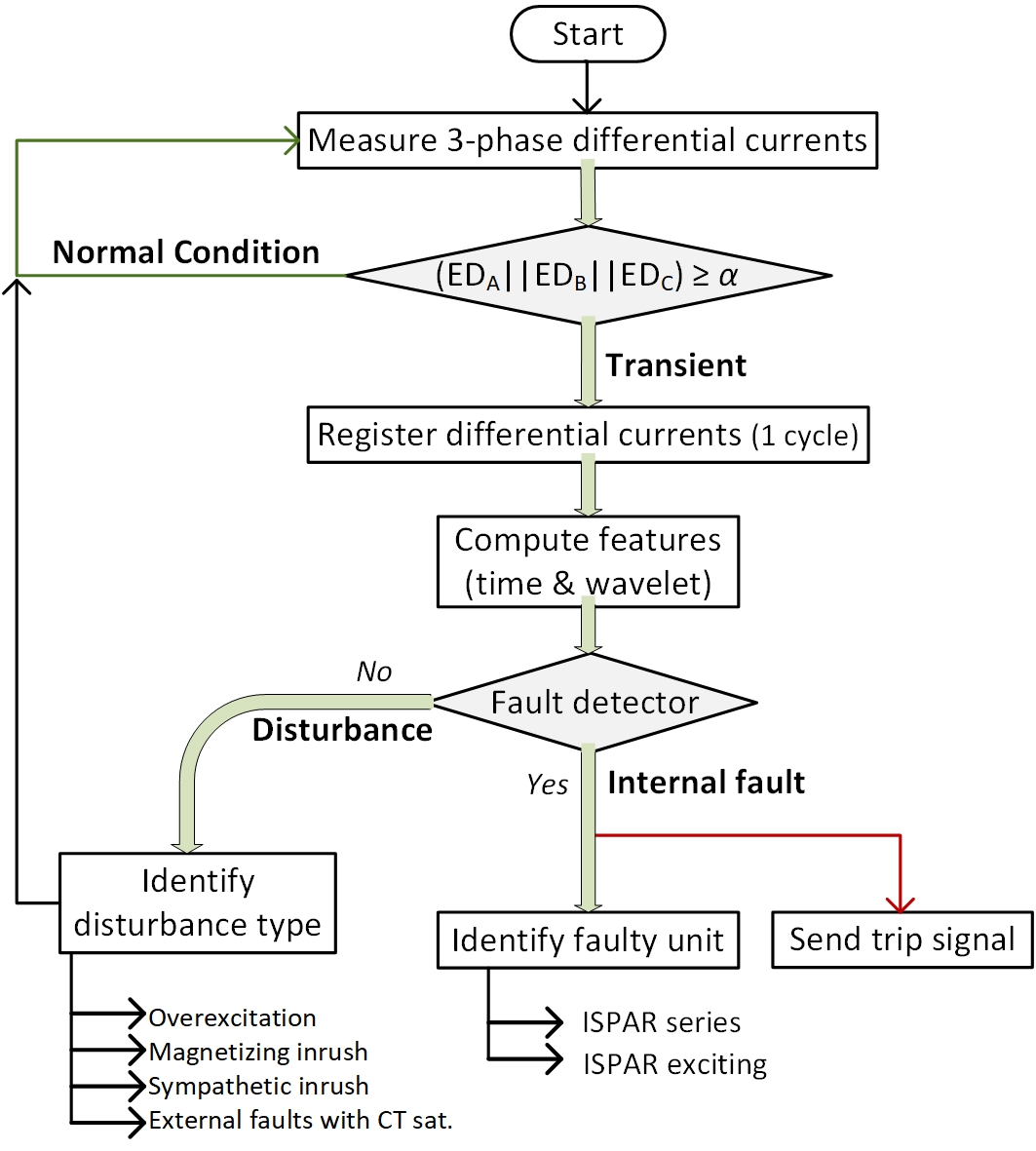}}
\vspace{0mm}\caption{Flowchart for fault detection and localization }
\label{flowchart}
\end{figure}

\section{{The proposed PAR Differential Protection}}
Figure \ref{flowchart} depicts the time and time-frequency domain based proposed protection and classification scheme having three applications: internal fault detection, fault localization, and identification of other transient disturbances.
The event detector {(ED)} detects the change in the differential currents if the ED index in any phase is more than a certain threshold, $\alpha$ and registers one cycle of post transient 3-phase differential currents. These currents are preprocessed to obtain detail wavelet coefficients, wavelet energy, and time-domain features. The decision on tripping is made by a classifier-based decision-making module. If the module detects an internal fault the relay operates and the faulty unit is identified otherwise the specific transient disturbance is identified.

\subsection{Event Detection}
The change in the differential currents in case of a transient event is detected by the ED which calculates the fractional increase in the difference between the cumulative sum of modulus of two consecutive cycles, given by:
\begin{equation} ED (t) = \frac{\sum_{i=t}^{n_c+t}|Id_\phi(i)|-\sum_{i=t}^{n_c+t}|Id_\phi(i-n_c)|}{\sum_{i=t}^{n_c+t}|Id_\phi(i)|}
\end{equation}
where i = sample number starting at the second cycle, $n_c$ = number of samples in one cycle, 
n = total number of samples, $Id$ = differential current, and $\phi$ = phase A,B, and C.
The event detection filter starts recording the data from the instant ED(t) $\geq \alpha$= 0.05 in any one of the 3-phases. In normal condition when there is no transient, the values of ED(t) are nearer to zero \cite{Dharmapandit2017}. These 3-phase differential currents are used to extract the essential features.

\subsection{Feature Selection Methods}\label{}
The success of any classification algorithm highly depends on the input features. Feature selection is critical in reducing the classification error. Given a dataset with \textit{M} features X=\{$x_i$, i=1,.., M\} and target variable \textit{y}, feature selection obtains from the M-dimensional space, a subspace of \textit{S} features that characterizes \textit{y}. Feature selection increases the interpretability, reduces time complexity, and improves the reliability of predictions.

\subsubsection{maximum Relevance Minimum Redundancy (mRMR)}
Feature selection methods based on mutual information, F-test select the top features without considering the relationship among the selected features. They calculate the mutual information as score between joint distribution of all features \textit{($x_i$)} and target \textit{y} and select the features with largest score. However, the selected features might be correlated and not cover the whole space. mRMR penalizes the relevancy of a feature obtained using the mutual information score by its redundancy when other features are also present. It searches for features, \textit{S} satisfying Eqn.\ref{maxrel} to select the features with highest mutual information \textit{I($x_i$;y)} to target variable \textit{y} and Eqn.\ref{minred} to reduce the redundancy of the features selected using maximum relevance (Eqn.\ref{maxrel}) \cite {mrmr}.
\begin{equation} \label{maxrel}
max D(S,y), D= \frac{1}{|S|}\sum_{x_i\in S}I(x_i;y)
\end{equation}
\begin{equation} \label{minred}
min R(S), R= \frac{1}{|S|^2}\sum_{x_i,x_j \in S}I(x_i,x_j)
\end{equation}
where \textit{I($x_i$;y)} and \textit{I($x_i$;$x_j$)} are mutual information which determine the amount of difference between the joint distribution and product of marginal distributions of the pair of random variables involved. 
\subsubsection {Random forest feature selection}\label{rfsel}
Random forest as a classifier performs implicit feature selection during training for classification which results in higher accuracy. This implicit feature selection is utilised to rank a feature $x_i$ by adding the impurity decrease $\Delta i(\tau,T)$ for all nodes $\tau$ where $x_i$ is used; averaged over all trees, T \cite{Breiman2001}.
\begin{equation} 
Imp(x_i)= \frac{1}{T}\sum_{T}\sum_\tau \Delta i(\tau,T) 
\end{equation}
where $i(\tau)$ is the `gini impurity' at node $\tau$, given by : $i(\tau) = 1-\sum_{i}^{c}(p_i|\tau)^2$. $p_i$ is the fraction of the samples that belong to class $c$. 

 The analysis of transient events depends on the effective processing of the 1-D differential signals. The features for the seven classifiers are obtained considering firstly, the time-domain features, secondly, the time-frequency domain features, and lastly, by combining the above two domains.

\subsection{Features Selected}
The composition of a signal can be analyzed by the different frequency components and time-domain statistics. Wavelet transform is suitable for decomposing an aperiodic signal into frequency bands and their time-frequency analysis has been used in several applications that require time and frequency information simultaneously: gait analysis, fault detection, ultra-wideband wireless communications, etc. 

\subsubsection{Differential Wavelet Coefficients} \label{wfeatures}
Discrete Wavelet Transform (DWT) quantifies the similarity between the original signal and the wavelet function by the detail ($d_l$) and approximation ($a_l$) coefficients \cite{WAVELET}. The low and high-frequency components are obtained at each decomposition level $l$ using : 
\begin {equation}
\footnotesize
\!\ \ a_l(k)\!\!=\!\!\!\sum_{l_k} w_\varphi(l_k\!-\!2k)a_{l-1}(l_k)\   \     
\    \  \ \ \ \  d_l(k)\!\!=\!\!\!\sum_{l_k}  w_\psi(l_k\!-\!2k)a_{l-1}(l_k)
\end {equation}
where $w_\varphi$, $w_\psi$ are the low and high pass filters. The mother wavelet and decomposition level used influences the detail coefficients and thus the classification accuracy. However, researchers \cite{flttransient_wave_energy,SVM1,SVM2,ann1,dtwt} have arbitrarily chosen the wavelet function and decomposition level without justifying their use.

Here, multilevel 1D DWT is used with wavelet families: `Daubechies', `Symlets', `Coiflets', `Biorthogonal', `Reverse biorthogonal', and `Discrete Meyer' to extract the wavelet coefficients. The wavelet functions in each wavelet family (`Daubechies'- db1 to db38, `Symlets'- sym2 to sym20, `Coiflets'- coif1 to coif14, `Biorthogonal'- bior1.1 to bior6.8, `Reverse biorthogonal'- rbio1.1 to rbio6.8, `Discrete Meyer'- dmey) are decomposed at different levels.  The maximum useful level of decomposition chosen to avoid edge effects caused by signal extension is given by:
Max level=$\lfloor  log_{2}(\frac{signal\ length}{filter\ length-1})\rfloor$.
Features (wavelet functions + level) for the {three different} tasks are chosen using a classifier-involved method. 
The detail coefficients of the 3-phase differential currents obtained from each of these wavelet functions at permissible decomposition levels are used to train and test Decision Trees (DT) finding the one which minimizes the error rate. Five wavelet coefficients with the best-balanced accuracies averaged over 5 runs are selected. 
bior2.2 at 3, db4 at 4, rbio3.3 at 3, rbio4.4 at 4, and sym4 at 4  are the top 5 performers for the detection of internal faults. bior1.3 at 1,  db1 at 1, coif1 at 3, rbio1.1 at 4, and sym2 at 4 are for locating the faulty units. bior4.4 at 4, bior2.2 at 5, db2 at 5, rbio1.1 at 3, and sym2 at 5 are for identification of transients.
Once the top 5 wavelet functions and the corresponding decomposition levels are obtained using the DT (the base classifier here), the wavelet coefficients are used to train and test six other well known classifiers.

\subsubsection{Time-Domain Features}\label{td features}
A comprehensive number of time-domain features are also extracted from the 3-phase differential currents obtained from the current transformers (CTs). The complete list of the features extracted can be found in \cite{tsfresh}. Random Forest is used to rank and select those time-domain features having maximum information gain to distinguish between the different classes. The most relevant features for detection, localization and classification tasks obtained after performing feature ranking belong to the set \textbf{F} = \{F1, F2, F3, F4, F5, F6, F7\}  where, F1 = average change quantile, F2 = maximum, F3 = autoregressive coefficients, F4 = aggregate linear trend,  F5 = autocorrelation, F6= number of peaks, F7= energy ratio by chunks. 
(F1, F2, F3), (F1, F3, F4), and (F5, F6, F7) are chosen for the detection of internal faults, localization of faulty units, and identification of other transients respectively. These time-domain features are detailed in the following part. 
\begin{itemize}

\item[$\square$] F1, autoregressive coefficients are the least-square estimates of $\varphi_{i's}$ which are obtained by minimizing Eqn.\ref{ar} with respect to $\varphi_0, \varphi_1..., \varphi_l$ and time lag $l$.
\begin{equation}\label{ar} \sum_{t=l+1}^{n_c} [Id_{\phi_t}- \varphi_0 - \varphi_1\cdot Id_{\phi_{t-1}} - ...-\varphi_l \cdot Id_{\phi_{t-l}}]^2 \end{equation}

\item[$\small\square$] F2, maximum, calculates the maximum value in the signal, $Id_\phi$.

\item[$\small\square$] F3, average change quantile, calculates the average of absolute values of consecutive changes of the time series inside two constant values $qh$ and $ql$ with $n'$ number of sample points.
\begin{equation} avg.\  change\ quantile =\frac{1}{n'}\cdot{\sum_{t=1}^{n'-1} |Id_{\phi_{t+1}} - Id_{\phi_{t}}| }
\end{equation} 

\item[$\square$]F4, aggregate linear trend, calculates the linear least-squares regression for values of the time series over windows and returns aggregated values of either intercept or standard error. 

\item[$\square$]F5, autocorrelation, calculates similarity between observations of the signal with variance, $\sigma^2$ and mean, $\mu$ as a function of time lag $l$.
\begin{equation} autocorrelation=\frac{1}{(n_c-l)\sigma^2}{\sum_{t=1}^{n_c-l} (Id_{\phi_t}-\mu) (Id_{\phi_{t+l}}-\mu) } 
\end{equation}

\item[$\square$] F6, number of peaks, calculates the number of peaks with minimum support m in the differential currents.

\item[$\square$] F7, energy ratio, calculates the ratio of energy of jth chunk and the entire signal, $Id_\phi$ with chunk length $n'$.
\begin {equation}
\mathcal{E}_r={\sum_{t=1}^{n'}Id_{\phi_{jt}}^2}\Big/ {\sum_{t=1}^{n}Id_{\phi_t}^2}
\end {equation} 
\end{itemize}

\subsubsection{Differential Wavelet Energy \& Time-Domain Features}
The time-domain features selected using maximum information gain in Section \ref{td features} and detail wavelet coefficients energy are combined to form a new set of inputs. At first, the energy associated with the wavelet coefficients for each wavelet function at all permissible decomposition levels considering one cycle post transient 3-phase differential is calculated using Eqn.\ref{we}.
\begin {equation}\label{we}
\mathcal{E}_{dl}^w = \sum_{k}|d_l(k)|^2 
\end {equation}
Then nine most relevant wavelet energy features are obtained using the maximum Relevance Minimum Redundancy (mRMR) algorithm. mRMR finds the optimal subset of features by considering both the features importances and the correlations between them. 18 features ( time-domain(9) + wavelet energy(9)) are used to train the seven classifiers for the detection, localization and classification tasks.

\subsection{ Classifiers Used}
 Different classifiers are used to evaluate the validity of the proposed feature-based protection scheme. Tree-based learning algorithms: decision trees (DT), random forest (RF), and gradient boosting (GB) are considered among the predominantly used supervised learning methods in problems related to data science. They have higher accuracy and are easy to interpret. The other classifiers are Naive Bayes (NB)- a probabilistic classifier competitive in certain domains; Support-vector machines (SVM)- basically a non-probabilistic classifier that maps the inputs in high dimensional space; Neural Networks (NN)- inspired by the human brain and adapted in a variety of applications ranging from computer vision, social networking to painting over time; and k-nearest neighbors (kNN), lazy learners, where the system generalizes the training data after receiving a query, in contrast to the other classifiers used.
\subsubsection{Decision Tree}
DTs are distribution-free white box Machine Learning models that learn simple decision rules inferred from the feature values to predict the target. The CART algorithm introduced by Breiman et al \cite{Breiman} and implemented in scikit-learn is used which constructs binary trees by splitting the training set recursively till it reaches the maximum depth or a splitting doesn't reduce the impurity measure (`gini', `classification error', or `entropy'). The candidate parent node is split into child nodes using a feature ($x_i$) and threshold that yields the largest information gain. DT has also been used as the base classifier to select the wavelet coefficients in Section \ref{wfeatures}.

\subsubsection{Random Forest}
RF belongs to the family of ensemble trees which builds numerous base estimators and averages their predictions producing a better estimator with reduced variance. Each tree constitutes a random sample (drawn with replacement) of the training set and the best split is found at each node by considering a subset of input features. The individual trees tend to overfit but averaging the predictions of all trees reduces the variance \cite{Breiman2001}. The hyperparameters in RF: number of trees, tree depth, and feature size to consider when splitting is optimized using the hyperparameter search. RF is also used during feature selection and ranking (Section \ref{rfsel}) to get the relative importance of the features which is measured by the fraction of samples a feature contributes to and the mean decrease in impurity from splitting the samples \cite{phdthesis}.


\subsubsection{Gradient Boosting}
GB is also a class of ensemble trees which builds the base estimators from weak learners sequentially in a greedy manner resulting in a strong estimator \cite{friedman2001} \cite{Mason}. The newly added learner tries to minimize the loss function.
GB uses shrinkage which scales the contribution of the weak learners by the learning rate and sub-sampling of the training data (stochastic gradient boosting) for regularization. The main hyperparameters in GB -- no of estimators, max depth, and learning rate -- are obtained using grid search.

\subsubsection{Naive Bayes} Based on the Bayes rule, NB \cite{nb} is one of the oldest and simplest classifiers which assumes that feature variables are independent of each other given the target variable.
NB has been shown to have good classification performance for many applications such as text categorization, medical diagnosis.

\subsubsection{Support Vector Machines}
SVM \cite{svmc1,svmc2} are memory-efficient modern classifiers which uses kernels to construct hyperplanes or linear classification boundaries in higher dimensional spaces. Different kernels: `rbf', `poly', and `sigmoid' with different gamma and regularisation parameter, C values are used during the hyperparameter optimization.

\subsubsection{Neural Network}
The Neural Network  (NN) \cite{mlpbook} consists of one or more hidden layers (composed of a number of perceptrons) which learn the approximation function $f(.): R^S \xrightarrow{}R^c$ with $S$ number of features and $c$ number of outputs. Hyperparameters: the hidden layer size, activation function, regularization term, learning rate, are optimized using grid search.

\subsubsection{k-Nearest Neighbor}
Nearest Neighbor is an instance based non-parametric learning which computes the class of an instance by majority voting of the k (an integer) nearest neighbours of each query point. The training phase involves storing the features and target labels. Distance measure: L1 (Manhattan) and L2 (Euclidean), number of neighbours are optimised to get the best hyperparameters \cite{knn}.
\begin{table}[ht]
\caption{Performance with wavelet coefficients for internal fault detection}\label{wfaults}
\setlength{\tabcolsep}{2.2pt}
\begin{tabular}{@{}ccccccccc@{}}
\toprule
\multicolumn{1}{c}{\multirow{2}{*}{\begin{tabular}[c]{@{}c@{}}  \textit{Wavelet}\\  \textit{Function}\end{tabular}}} & \multicolumn{1}{c}{\multirow{2}{*}{\begin{tabular}[c]{@{}c@{}} \textit{Decomposition}\\   \textit{Level}\end{tabular}}} & \multicolumn{7}{c}{\textit{Classifier($\bar{\eta}$)}} \\ \cmidrule(l){3-9} 
\multicolumn{1}{c}{}                                                                            & \multicolumn{1}{c}{}                                                                               &  DT & RF & GB& NN  & kNN &  NB &  SVM \\ \midrule
bior2.2 & 3 & 97.4   & 99.6 & 99.7 & 99.8 &  99.5 &   71.0  & 90.2\\
db4   &  4   & 97.4   & 99.6   & 99.7        & 99.6        & 99.5          & 77.1           & 93.0\\
rbio3.3   & 3    &  97.6   & 99.7   & 99.9   & 99.6   & 99.2   & 76.5    & 93.0\\
rbio4.4   & 4    &   97.7 &  99.8  &  99.8  &99.8 & 99.8   &  76.2  &  97.0  \\
sym4 & 4 & 97.6 & 99.8 & 99.8 & 99.7 & 99.6 & 77.7&  93.8\\  \hline
\end{tabular}
\end{table}

\begin{table}[ht]
\caption{Performance with wavelet coefficients for locating the faulty unit}\label{wloc}
\setlength{\tabcolsep}{2.2pt}
\begin{tabular}{@{}ccccccccc@{}}
\toprule
\multicolumn{1}{c}{\multirow{2}{*}{\begin{tabular}[c]{@{}c@{}}  \textit{Wavelet}\\  \textit{Function}\end{tabular}}} & \multicolumn{1}{c}{\multirow{2}{*}{\begin{tabular}[c]{@{}c@{}} \textit{Decomposition}\\  \textit{Level}\end{tabular}}} & \multicolumn{7}{c}{\textit{Classifier($\bar{\eta}$)}} \\ \cmidrule(l){3-9} 
\multicolumn{1}{c}{}                                                                            & \multicolumn{1}{c}{}                                                                               &  DT & RF & GB& NN  & kNN &  NB &  SVM \\ \midrule
bior1.3 & 1 &  93.2 & 95.7 & 96.5 &93.3 & 94.1 &  56.1& 89.2\\

coif1 & 3 & 93.1  & 94.2 & 97.3 &  93.3 &94.7  &60.7&89.3 \\
db1 & 1 & 92.7   & 95.2 & 96.5 & 94.3 & 94.0  & 56.1 & 88.8  \\
rbio1.1 & 4 & 92.7  &  95.1&  96.1& 91.0 &93.5& 52.4 & 87.2\\
sym2 & 4 & 92.7  & 93.9 & 96.8 & 93.5 &  93.9& 56.6 & 89.4 \\  \hline
\end{tabular}
\end{table}


\begin{table}[ht]
\caption{Performance with wavelet coefficients for classification of transients}\label{wtransients}
\setlength{\tabcolsep}{2.2pt}
\begin{tabular}{@{}ccccccccc@{}}
\toprule
\multicolumn{1}{c}{\multirow{2}{*}{\begin{tabular}[c]{@{}c@{}}  \textit{Wavelet}\\  \textit{Function}\end{tabular}}} & \multicolumn{1}{c}{\multirow{2}{*}{\begin{tabular}[c]{@{}c@{}} \textit{Decomposition}\\   \textit{Level}\end{tabular}}} & \multicolumn{7}{c}{\textit{Classifier($\bar{\eta}$)}} \\ \cmidrule(l){3-9} 
\multicolumn{1}{c}{}                                                                            & \multicolumn{1}{c}{}                                                                               &  DT & RF& GB &  NN  & kNN &  NB &  SVM \\ \midrule
bior2.2 & 5 & 95.9  &98.1 & 98.0& 99.2 & 99.4  & 73.7 & 97.9 \\
bior4.4 & 4 & 96.1  & 98.0& 99.2& 99.5 & 98.2 & 70.9  & 97.6\\
db2 & 5 & 95.6  & 96.1& 99.4& 98.7 &  98.8& 72.2 & 98.0\\
rbio1.1 & 3 & 95.8  & 97.7& 98.6& 98.8 & 98.9 & 74.9 & 98.1\\
sym2 & 5 &  96.1 & 96.2& 97.9  & 99.5 &  99.5&  71.3 &   97.8  \\ \hline
\end{tabular}
\end{table}


\begin{table}[ht]
\footnotesize
\renewcommand{\arraystretch}{0.75}
\setlength{\tabcolsep}{2.4pt}
\caption{Performance with time-domain features}\label{tall}
\begin{tabular}{@{}lcccccccc@{}}
\toprule
\multirow{2}{*}{\begin{tabular}[c]{@{}l@{}}
\footnotesize{\textit{Application} }
\end{tabular}} & \multirow{2}{*}{\textit{Features}} &      & \multicolumn{5}{c}{\textit{Classifiers($\bar{\eta}$)} }\\ \cmidrule(l){3-9} 
                                                                                                       &                           & DT   & RF  & GB    & NN      & kNN       & NB       & SVM      \\\addlinespace[0.0cm]  \midrule
                                                                                                      
\textit{detect faults }                                                                                       & F1,F2,F3    & 97.0 &    96.3     &    99.3   & 94.6 & 93.3      & 77.8     & 90.7     \\ \addlinespace[0.1cm] 
 \textit{locate faults}                                        & F1,F3,F4    & 94.7 &     94.8     &    97.8  & 84.0  & 90.5      & 60.0       & 82.1     \\ \addlinespace[0.1cm] 
\textit{identify transients }                                & F5,F6,F7    & 95.4 &      98.3 &   98.5    & 97.2 & 97.2      & 76.9     & 97.1     \\ \bottomrule
\end{tabular}
\end{table}


\begin{table}[ht]
\caption{Performance with time-domain and wavelet features }\label{tw}
\setlength{\tabcolsep}{3pt}
\begin{tabular}{@{}llllllll@{}}
\toprule
\multicolumn{1}{l}{\multirow{2}{*}{\begin{tabular}[c]{@{}c@{}} \textit{Application}\end{tabular}}} & \multicolumn{7}{c}{\textit{Classifier($\bar{\eta}$)} }\\ \cmidrule(l){2-8} 
\multicolumn{1}{c}{}                                                                            & \multicolumn{1}{c}{}                                                                                DT & RF & GB & NN  & KNN &  NB &  SVM \\ \midrule
    \textit{detect faults}& 97.3 &99.2  &99.8& 94.7 & 95.6 & 78.0 &93.4\\
 \textit{locate faults} & 94.9  & 98.2 & 98.7 & 90.1  & 93.7  & 62.0 & 87.6\\
  \textit{identify transients} &  97.0 & 98.6  & 99.3 & 99.2 & 99.0 &64.9 &99.1\\
 \hline
\end{tabular}
\end{table}

\section{Results and Discussions}
At a sampling frequency of 10kHz, 167 samples per cycle per phase are available for further processing. The features extracted and selected from the one cycle of post transient 3-phase differential currents obtained from the CTs and registered by the ED is used to train the different classifiers. The input dimension of the training and testing cases varies depending on the level of decomposition and wavelet function chosen when wavelet coefficients are used to extract the features. In the case of time-domain features, the input dimension is 9 and when time and wavelet features are combined to train the classifiers the dimension is 18. To reduce the classification error 10-fold stratified cross validation and grid search are applied which use the available data effectively and train the classifiers on optimized hyperparameters. Normally, accuracy is used as the typical metrics to measure the performance of a classifier. But, it gives biased results if the data is imbalanced. Since, the classes are imbalanced in all the three tasks, balanced accuracy which is the average accuracy obtained on all classes and computed as $\bar{\eta}$ = $ \frac{1}{2}\cdot[\frac{TP}{(TP+FN)} + \frac{TN}{(TN+ FP)}]$ for binary classes is used to compute the performance where, TP = true positive, TN = true negative, FP = false positive, and FN = false negative \cite{imbalance}.

\subsection{Detection of internal faults}
Since the occurrence of any power system transient event is unpredictable in time, the use of an ED becomes imperative. Out of the {three mentioned} classification tasks, correct discrimination of internal faults from the other transients is the foremost. The dependability and security of the proposed method depend on the FP and FN of this binary class problem. The less the classification error, the better is the performance of the entire scheme. To achieve this the seven well-known classifiers are trained on 48,442 cases and tested on 12,110 cases which are simulated in section \ref{sec2}.
The seven classifiers are trained with three sets of features and the testing accuracies are reported. First, the selected wavelet coefficients obtained using exhaustive search by training DT are used as the inputs and the classification performance is shown in Table \ref{wfaults}. RF, GB, NN, and kNN performed well with GB giving $\bar{\eta}$ = 99.9\% on `rbio3.3' at level 3.  Second, the time-domain features F1, F2, and F3 of the 3-phase currents are used. Table \ref{tall} shows the balanced classification accuracy of the seven classifiers. GB gives the best performance with $\bar{\eta}$ = 99.3\%. Third, the classifiers are trained on the nine features from F1, F2, and F3 and the top nine wavelet energy features obtained using the mRMR algorithm. Table \ref{tw} shows the classification performance on the 18 features (time-domain + wavelet energy) of the seven different classifiers. GB and RF performed well with GB having an accuracy of $\bar{\eta}$ = 99.8\%. Table \ref{tabflt} shows the best results for detection of internal faults obtained with wavelet coefficients and GB on hyperparameters : `learning rate'= 0.1, `no of estimators'= 5000, `max depth'= 5.

\subsection{Locate faulty unit}
After the fault detection scheme detects an internal fault, the faulty transformer unit (ISPAR Exciting or ISPAR Series) is identified using the one-cycle of the post fault differential currents. To locate the faulty transformer unit 37, 498 fault cases are trained and 9,374 cases are tested. Table \ref{wloc} shows the classification performance on selected wavelet coefficients as features using seven different classifiers and Table \ref{tall} shows the same for time-domain features. Best performances of $\bar{\eta}$ = 97.3\% is obtained on `coif1' at level 3 and $\bar{\eta}$ = 97.8\% on F1, F3, and F4 3-phase differential time-domain features with GB. Next, the nine time-domain features from F1, F3, and F4 and the top nine time-frequency features obtained from the wavelet energy of the coefficients using mRMR are used to train the classifiers again. Table \ref{tw} shows the classification performance on the 18 features of the seven different classifiers. GB gives an improved accuracy of $\bar{\eta}$ = 98.7\%.
Table \ref{tabloc} shows the best results for identification of faulty unit obtained with F1, F3, F4 and wavelet energy as inputs and GB as the classifier on 
hyperparameters : `learning rate'= 0.05, `max depth'= 10, `no of estimators'= 12000.

\subsection{Classification of transient disturbances}
The various disturbances: magnetizing inrush, sympathetic inrush, overexcitation, and external faults with CT saturation are classified using one-cycle of post-transient samples after they are differentiated as no-fault by the fault detection scheme. Table \ref{wtransients} shows the classification performance on selected wavelet coefficients as features using seven different classifiers and Table \ref{tall} shows the same for time-domain features. The classifiers are trained on 10,994 cases and tested on 2,736 cases. Best performances of $\bar{\eta}$ = 98.5\% are obtained with GB on F5, F6, and F7 time-domain features. $\bar{\eta}$ of 99.5\%  with NN on `bior4.4' at level 4 and `sym2' at level 5 and with kNN on `sym2' at level 5 are obtained with the wavelet coefficients.
GB, NN, kNN, and SVM showed satisfactory results with GB having the least error with $\bar{\eta}$ = 99.3\% (table \ref{tw}) when trained with 18 features (time-domain + wavelet energy). Table \ref{tabdisturbance1} shows the best results for identification of transient disturbances obtained with wavelet coefficients and NN on hyperparameters : `learning rate'= `adaptive', `solver'= 'adam', `activation'= 'relu', `alpha'= 0.01, `hidden layer size' = ( 51, 13).

\begin{table}[ht]
\setlength{\tabcolsep}{1.7 pt}

\caption{Detection of faults \& Localization of faulty unit }
\begin{subtable}[l]{.53\linewidth}
\caption{Detect internal fault } \label{tabflt}
{\begin{tabular}{lllll} \toprule
\textit{Fault / no-fault}  & Total & TP  & FN & FP \\ \midrule
\textit{Internal faults}   & 9406 & 9406 & 0 & 8 \\
\textit{Disturbances}      & 2705 & 2697 & 8 &0    \\ \hline

\end{tabular}}{}

\end{subtable}\hfill
\begin{subtable}[r]{.43\linewidth}
\caption{Locate faulty unit} \label{tabloc}
{\begin{tabular}{lllll}\toprule
\textit{Unit } & Total & TP  & FN & FP \\ \midrule
\textit{Series}   & 6707 & 6695 & 12 & 67\\
\textit{Exciting } & 2668 & 2601 & 67 & 12\\ \hline
\end{tabular}}{}
\end{subtable}
\end{table}
\vspace{-3mm}

\begin{table}[ht]
\centering
\caption{Identification of no-fault transient disturbances\label{tabdisturbance1}}
{\begin{tabular}{lllll}\toprule
\textit{Transient Disturbances }  & Total & TP  & FN &   FP\\
\midrule
\textit{Magnetizing Inrush} &  489 & 486 & 3 & 3\\
\textit{Sympathetic Inrush } & 552 & 546 & 6 & 4\\
\textit{Overexcitation} & 156 & 156 & 0 & 2\\
\textit{External faults with CT sat.} & 1539 & 1535 & 4 & 4\\
\hline
\end{tabular}}{}
\end{table}
Although the most informative features from a high-dimensional feature set and from time, time-frequency, and their combinations were used to train different classifiers to perform the three classification tasks, it was observed that there is no specific set of features and classifier model alliance which performs reliably in all the three applications and outperforms the others. Nevertheless, several trends that are visible are:\begin{itemize}
\vspace{-0.3cm}
\item[$\square$]
With wavelet coefficients as feature RF, GB, NN, kNN had fewer misclassifications when `discriminating faults from other transients'; while `locating faulty unit' the best balanced accuracy was 97.3\% given by GB which was improved when time and wavelet energy features were combined to 98.7\%; and while `identifying the transients' GB, NN, and kNN performed well. 
\item[$\square$] With time-domain features GB differentiates between the classes in all three applications better than the other classifiers used.
\item[$\square$] With time and wavelet features GB performed better, however, kNN, NN and SVM also had lower classification error in case of `identifying the transients'.
\end{itemize}
The execution time for the feature extraction, training, and testing of the best performing classifiers for each task using one CPU core are reported using the in-built library in python. The time is averaged over 100 runs. The execution time for the different sections of the algorithm for the three tasks is shown in Table \ref{ptime}. The testing time column shows both the execution time for one test instance and all test instances. The in-service computations in the fault/no-fault decision will include only testing a single instance and feature calculation. The columns `Testing time One' and `Feature extract time' show that the proposed scheme takes an additional 0.925 ms after one cycle of inception of fault to detect a fault. Locating the faulty unit and identifying the transient will take 8.45 ms and 263 $\mu$s respectively. Noting that computations can be further optimized, these processing times are suitable for future real-time implementation.

\begin{table}[ht!]
\caption{Execution time for the best classifiers}\label{ptime}
\setlength{\tabcolsep}{1pt}

\begin{tabular}{@{}lccccc@{}}
\toprule
\multirow{2}{*}{\textit{Application}} & \multicolumn{1}{c}{\multirow{2}{*}{\begin{tabular}[c]{@{}c@{}} \textit{Classifier}\end{tabular}}} & \multicolumn{1}{c}{\multirow{2}{*}{\begin{tabular}[c]{@{}c@{}} \textit{Training}\\\textit{time}\end{tabular}}} & \multicolumn{2}{c}{\textit{Testing time}} & \multicolumn{1}{c}{\multirow{2}{*}{\begin{tabular}[c]{@{}c@{}}\textit{Feature extract}\\\textit{time}\end{tabular}}} \\ \cmidrule(lr){4-5}
                             & \multicolumn{1}{c}{}                                                                           & \multicolumn{1}{c}{}                                                                         & One          & All          & \multicolumn{1}{c}{}                                                                                   \\ \midrule
\textit{detect faults  }              &   GB             &       697s      &       0.8ms        &    0.38s            &      125  $\mu s $       \\
\textit{locate faulty unit }          &  GB    &    408s     &    2.7ms            &      0.53s              &   5.75 ms             \\
\textit{identify transients  }        & NN   &    5.6s    &     76$\mu s $     &  2.5ms   &    187  $\mu s $    \\ \midrule
\end{tabular}
\end{table}

The ED is implemented in MATLAB 2019b while the classifiers are built-in Python 3.8 using Scikit-learn framework \cite{scikit2}. The no\_of\_jobs parameter was also used to parallelize the computation of predictions by using more processing units. The pre-processing of the 3-phase currents is done in Python and MATLAB. The PSCAD simulations are carried out on Intel Core i7-6560U CPU @ 2.20 GHz, 8 GB RAM and the classifiers are run on Intel Core i7-8700 CPU @ 3.20 GHz, 64 GB RAM.

\section{Conclusion}\label{sec5}
This paper addresses the problem of detection and localization of internal faults and the identification of other transient disturbances for an ISPAR. The internal faults are distinguished from overexcitation, inrush, and external faults with CT saturation conditions. Then the faulty unit of the ISPAR (series or exciting) is located or the transient disturbance is identified depending on whether an internal fault was detected. The event detector detects the variation in differential current and registers one cycle of 3-phase post transient samples which are then used to extract time and time-frequency features and train seven different classifiers. Firstly, the classifiers are trained with the most important wavelet coefficients (wavelet function + level) obtained by exhaustive search using Decision Tree. Secondly, time-domain features selected by maximizing the information gain is used to train these classifiers. Lastly, the top nine wavelet features (energy of wavelet coefficients) obtained using maximum Relevance Minimum Redundancy and time-domain features are used for training. Although it is observed that Gradient Boost Classifier performed satisfactorily across the board, the choice of features -- wavelet coefficient or time–domain or wavelet energy and time-domain -- is application dependent. The proposed scheme can work together with a traditional microprocessor-based differential relay offering supervisory control over it's operation and thus improving security and dependability. PSCAD/EMTDC is used for simulation of the transient events and to model the ISPAR. 

\section{Appendix}\label{sec14}
Fortran code for single-phase 3-winding transformer fault model
\vspace{-4mm}
\begin{table}[H]
\footnotesize
\setlength{\tabcolsep}{2pt}
\renewcommand{\arraystretch}{0.75}
 \begin{tabular}{ll}\toprule
  1. nw = 6, w = 2*pi*f         &         23. L3m = v2/(w*$I_m2$*i2)*$fc^2$\\
  2. $I_m3$ = $I_m2$ =$I_m1$    &           24. L4m = v2/(w*$I_m2$*i2)*$fd^2$ \\
  3. fa = fault1*0.01        &       25.  L5m =v3/(w*$I_m3$*i3)*$fe^2$ \\
  4. fb = 1.0-fa        &       26. L6m =v3/(w*$I_m3$*i3)*$ff^2$ \\
  5. fc = fault2*0.01    &       25. L1 = L1l + L1m, L2 = L2l + L2m \\
  6. fd = 1.0-fc          &        26. L3 = L3l + L3m, L4 = L4l + L4m  \\
  7. fe = fault1*0.01     &  27. L5 = L5l + L5m, L6 = L6l + L6m  \\
  8. ff = 1.0-fe           &   28. M12 = sqrt(L1m*L2m)  \\
  7. z1=v1/i1, z2=v2/i2, z3=v3/i3 &        29. M13 = sqrt(L1m*L3m)\\
  9. l1 = v1/(w*$I_m1$*i1)              &       30. M14 = sqrt(L1m*L4m)\\
  10. l2 = v2/(w*$I_m2$*i2)              &        31.  M15 = sqrt(L1m*L5m)\\
  11. l3 = v3/(w*$I_m3$*i3)          &   32. M16 = sqrt(L1m*L6m)\\
  12. X1 = (x13-x23+x12)/2    & 34. M23 = sqrt(L2m*L3m)\\
  13. X2 = (x23-x13+x12)/2   & 35. M24 = sqrt(L2m*L4m)\\
  14. X3 = (x13-x12+x23)/2    & 36. M25 = sqrt(L2m*L5m)\\
  15. Lk1 = X1*z1/w &             37.  M26 = sqrt(L2m*L6m)\\
  16. Lk2 = X2*z2/w &           38. M34 = sqrt(L3m*L4m)\\
  17. Lk3 = X3*z3/w &            39. M35 = sqrt(L3m*L5m)\\
  18. L1l= Lk1*$fa$, L2l=Lk1*$fb$   &          40. M36 = sqrt(L3m*L6m)\\
  19. L3l= Lk2*$fc$, L4l=Lk2*$fd$   &         41. M45 = sqrt(L4m*L5m)\\
  20. L5l= Lk3*$fe$, L6l=Lk3*$ff$   & 42. M46 = sqrt(L4m*L6m)\\
  21. L1m =v1/(w*$I_m1$*i1)*$fa^2$  & 43.  M56 = sqrt(L5m*L6m)\\ 
  22. L2m =v1/(w*$I_m1$*i1)*$fb^2$   &                         \\ \hline
\end{tabular}
\end{table}
\bibliographystyle{iet.bst}
\balance
\bibliography{references}
\end{document}